\documentclass[%
 prb,%
 amsmath,amssymb,
preprint,%
 showkey,
 noshowpacs,
 superscriptaddress
]{revtex4-1}

\usepackage{graphicx}
\usepackage{dcolumn}
\usepackage{bm}
\usepackage{chemformula} 
\usepackage{mathtools} 
\RequirePackage[normalem]{ulem}
\RequirePackage{color}\definecolor{RED}{rgb}{1,0,0}\definecolor{BLUE}{rgb}{0,0,1}

\begin{document}


\title{Perturbation approach to \textit{ab initio} effective mass calculations}

\author{Oleg Rubel}
\email[O.R. email: ]{rubelo@mcmaster.ca, ORCID: 0000-0001-5104-5602}
\affiliation{Department of Materials Science and Engineering, McMaster University, 1280 Main Street West, Hamilton, Ontario L8S 4L8, Canada}

\author{Fabien Tran}
\email[]{ORCID: 0000-0003-4673-1987}
\affiliation{Institute of Materials Chemistry, Vienna University of Technology, Getreidemarkt 9/165-TC, A-1060 Vienna, Austria}

\author{Xavier Rocquefelte}
\email[]{ORCID: 0000-0003-0191-2354}
\affiliation{Univ Rennes, CNRS, ISCR (Institut des Sciences Chimiques de Rennes) UMR
6226, F-35000 Rennes, France}

\author{Peter Blaha}
\email[P.B. email: ]{pblaha@theochem.tuwien.ac.at, ORCID: 0000-0001-5849-5788}
\affiliation{Institute of Materials Chemistry, Vienna University of Technology, Getreidemarkt 9/165-TC, A-1060 Vienna, Austria}

\date{\today}

\begin{abstract}
A degenerate perturbation $\bm{k}\cdot\bm{p}$ approach for effective mass calculations is implemented in the all-electron density functional theory (DFT) package WIEN2k. The accuracy is tested on major group IVA,  IIIA-VA, and IIB-VIA semiconductor materials. Then, the effective mass in graphene and CuI with defects is presented as illustrative applications. For states with significant Cu-$d$ character additional local orbitals with higher principal quantum numbers (more radial nodes) have to be added to the basis set in order to converge the results of the perturbation theory. Caveats related to a difference between velocity and momentum matrix elements are discussed in the context of application of the method to non-local potentials, such as Hartree-Fock/DFT hybrid functionals and DFT+$U$.
\end{abstract}

\maketitle

\section{Introduction}

Effective mass is one of the key concepts in the band theory of solids. It captures change in the energy dispersion relation for electrons as a result of their propagation in a periodic potential in contrast to the free space. In semiconductors, the effective mass at band edges has numerous implications. Those implications include the following: charge carrier transport coefficients, effective density of states, carrier concentrations and the position of the Fermi level at finite temperature, energy levels of shallow impurities, and the exciton binding energy (Ref.~\citenum{Ashcroft_Solid-State-Physics}, chaps.~28--30).  In metals, a nearly-free \textit{vs} heavy-fermion behaviour \cite{Stewart_RMP_56_1984} as well as a damping of the de Haas--van Alphen oscillations amplitude \cite{Shoenberg_Magnetic_1984} are intimately linked to band effective masses at the Fermi energy.

The importance of effective mass made it a wanted material characteristics extracted in post-processing of electronic structure calculations. The most straightforward way to determine the effective mass is to evaluate the curvature of the Bloch band
\begin{equation}\label{Eq:eff-mass-curvature}
	\left[ m^*_{\alpha\beta,n}(\bm{k}) \right]^{-1} = \hbar^{-2} \, \frac{\partial^2 E_n(\bm{k})}{\partial k_{\alpha}\partial k_{\beta}}
	,
\end{equation}
where $n$ is the band index, $\bm{k}$ is the electron wave vector,  $\alpha$ and $\beta$ are Cartesian directions ($x$, $y$, and $z$), and $\hbar$ is the reduced Plank's constant. Representative implementations of the finite difference [or $E_{n}(\bm{k})$ interpolative] technique include interpretation of thermoelectronic transport properties \cite{Madsen_CPC_175_2006,Mecholsky_PRB_89_2014} and de Haas–van Alphen frequencies \cite{Rourke_CPC_183_2012}.  Accurate determination of $m^*$ requires setting up  a finite difference grid in reciprocal space in the vicinity of the $k$ point of interest. The high-resolution $k$ point grid approach may not be practical for large supercells or computationally-intensive electronic structure methods that go beyond standard density functional theory (DFT)\cite{Hohenberg_PR_136_1964,Kohn_PR_140_1965}. One can of course resort to Wannier interpolation of the band structure \cite{Pizzi_JPCM_32_2020,w2wannier,Tillak_PRB_101_2020}, but it adds an additional layer of complexity. Thus, a Fourier interpolation of a band structure \cite{Madsen_CPC_175_2006} remains a method of choice for high-throughput studies of the effective mass \cite{Varley_CM_29_2017,Ricci_SD_4_2017}.

Alternatively, the $\bm{k}\cdot\bm{p}$ perturbation theory (PT) offers an elegant path to the calculation of effective masses that does not require a finite difference grid (see Ref.~\citenum{Ashcroft_Solid-State-Physics}, Appendix~E therein). The framework relies on a second order expansion of $E_n(\bm{k})$ in the vicinity of $\bm{k}_0$:
\begin{equation}\label{Eq:E(q)-second-order-q}
	E_n(\bm{k}_0 + \bm{q}) \approx
	E_n(\bm{k}_0 ) + 
	\sum_{\alpha} \frac{\partial E_n}{\partial k_{\alpha}} \, q_{\alpha} +
	\frac{1}{2} \sum_{\alpha,\beta} \frac{\partial^2 E_n}{\partial k_{\alpha} \partial k_{\beta}} \, q_{\alpha}q_{\beta}
\end{equation}
in terms of infinitely small $\bm{q} = \bm{k} - \bm{k}_0$. Energies $E_n(\bm{k}_0 + \bm{q})$ are eigenvalues of a $\bm{k}\cdot\bm{p}$ Hamiltonian  (with caveats discussed in Sec.~\ref{Sec:Results:Limitations}) for the cell-periodic wave functions
\begin{equation}\label{Eq:H(k0+q)-k.p}
	\hat{H}(\bm{k}_0 + \bm{q}) = 
	\hat{H}(\bm{k}_0) +
	\frac{\hbar}{m_0} \, \bm{q} \cdot \left( \hat{\bm{p}} + \hbar\bm{k}_0 \right) +
	\frac{\hbar^2}{2m_0} \, q^2
\end{equation}
whose matrix elements $H_{nm} = \langle u_{n\bm{k}_{0}} | \hat{H}(\bm{k}_0 + \bm{q}) | u_{m\bm{k}_{0}} \rangle$ are
\begin{equation}\label{Eq:Hnm-k.p}
	H_{nm} = 
	\begin{dcases}
    		E_n(\bm{k}_0) + \frac{\hbar}{m_0} \sum_\alpha \left[p_{nn\bm{k}_0}^{(\alpha)}+\hbar k_0^{(\alpha)}\right] q_{\alpha} + \frac{\hbar^2}{2m_0} \, q^2  & \quad \text{if } n=m\\
    		\frac{\hbar}{m_0} \sum_\alpha p_{nm\bm{k}_0}^{(\alpha)} q_{\alpha}  & \quad \text{otherwise}
  	\end{dcases}
\end{equation}
Here $m_0$ is the electron rest mass and $p_{nm\bm{k}_{0}}^{(\alpha)} = \left\langle u_{n\bm{k}_{0}} | \hat{p}_\alpha | u_{m\bm{k}_{0}} \right\rangle$ is the matrix element of the linear momentum operator $\hat{p}_{\alpha}=-i\hbar\,\partial/\partial r_{\alpha}$ with $u_{n\bm{k}}$ being the cell-periodic part of the Bloch wave function $\psi_{n\bm{k}}(\bm{r})=u_{n\bm{k}}(\bm{r}) e^{i\bm{k}\cdot \bm{r}}$.  The second order correction to $E_n(\bm{k}_0 + \bm{q})$ can be evaluated using PT, which yields the following expression for the effective mass (in non-degenerate case) \cite{Ashcroft_Solid-State-Physics}
\begin{equation}\label{Eq:eff-mass-nondegenerate}
	\frac{m_0}{m^*_{\alpha\beta,n}} = 
	\delta_{\alpha\beta} +
	\frac{1}{m_0}
	\sum_{l \neq n}
	\frac{
		p_{nl}^{(\alpha)}p_{ln}^{(\beta)} + p_{nl}^{(\beta)}p_{ln}^{(\alpha)}
	}{
		E_n - E_l
	}
	,
\end{equation}
where $\delta_{\alpha\beta}$ is the  Kronecker delta and the summation index $l$ runs over all occupied and empty bands (the $k$ point index is dropped for simplicity). Momentum matrix elements $p_{nl}^{(\alpha)}$ are readily available in modern electronic structure codes as a part of the linear optics and can be directly used in parametrization of the $\bm{k}\cdot\bm{p}$ Hamiltonian \cite{Willatzen_PRB_50_1994,Beresford_JAP_95_2004,Persson_CPC_177_2007,Lundie_JPCS_526_2014}. Thus, after an optic calculation is performed, the $k$- and band-resolved effective mass tensor $m^*_{\alpha\beta}$ can be obtained at a minimal computation cost. 

Notable achievements in adapting PT to \textit{ad initio} calculations of effective masses in solids include the work of \citet{Pickard_PRB_62_2000}. It represents  the first proof of concept for use DFT in conjunction with PT for calculation of effective masses in diamond while avoiding degeneracies. \citet{Shishidou_PRB_78_2008} adapted the method for use with linearized augmented plane waves allowing for degeneracies. In spite of promising developments, their proliferation into accessible DFT codes is very limited. To the best of our knowledge, \texttt{ABINIT} \cite{Gonze_CPC_248_2020} is the only \textit{ab initio} platform that offers DFT+PT calculation of effective masses thanks to the work of \citet{LaflammeJanssen_PRB_93_2016}.

Here we present the generalization of Eq.~(\ref{Eq:eff-mass-nondegenerate}) for degenerate bands and its implementation as an \texttt{mstar} code integrated into WIEN2k \cite{Blaha_WIEN2k_2018,Blaha_JCP_152_2020} DFT package. We verify the performance by calculating effective masses at the band extrema in Si, GaAs, and CdTe using the band curvature (calculated with numerical differentiation) \textit{vs} PT. To demonstrate capabilities of the new implementation, two illustrative examples are selected, namely, graphene and CuI (a system with valence $d$ electrons that present an additional challenge to the $\bm{k}\cdot\bm{p}$ theory \cite{Kuebbing_PRL_26_1971}).

\section{Method}

\subsection{Degenerate PT}

Degenerate states often appear at high-symmetry points of the Brillouin zone. Also, when spin-orbit coupling (SOC) is added, each band becomes at least double degenerate in structures with the inversion symmetry and without an external (or internal) magnetic field. Equation~(\ref{Eq:eff-mass-nondegenerate}) breaks down when $E_n=E_l$, \textit{i.e.}, two states belong to a subset of degenerate bands. This issue can be circumvented by using a degenerate PT (see Ref.~\citenum{Griffiths_Introduction-QM_2004}, chap.~6). In this case, evaluation of the effective mass is done in two steps.

Let us assume that we deal with a subset $D$ of degenerate bands in the range $[n \ldots m]$. The first step is the construction of a square matrix $M^{(\alpha\beta)}$ of the size $1+m-n$
\begin{equation}\label{Eq:Mnm}
	M_{nm}^{(\alpha\beta)} = 
	\frac{1}{m_0}
	\sum_{l\notin D}
	\frac{
		p_{nl}^{(\alpha)}p_{lm}^{(\beta)} + p_{nl}^{(\beta)}p_{lm}^{(\alpha)}
	}{
		E_D - E_l
	}
	.
\end{equation}
Here the index $l$ runs over all bands (occupied and unoccupied states, excluding core states) same as in Eq.~(\ref{Eq:eff-mass-nondegenerate}). The number of unoccupied bands is an important convergence parameter, in contrast to total energy calculations. The sensible energy range for empty bands (at least for $sp$ bonding) is about 5~Ry above the Fermi energy. The second step involves calculation of the effective mass for bands $n \ldots m$ using eigenvalues of $M^{(\alpha\beta)}$
\begin{equation}\label{Eq:eff-mass-degenerate}
	\frac{m_0}{m^*_{\alpha\beta,n \ldots m}} =
	\delta_{\alpha\beta} + \text{eig}
	\left[
		M^{(\alpha\beta)}
	\right]
	.
\end{equation}
One can see that Eqs.~(\ref{Eq:Mnm}) and (\ref{Eq:eff-mass-degenerate}) reduce to Eq.~(\ref{Eq:eff-mass-nondegenerate}) when $n=m$.

In practice, numerical inaccuracies due to a floating-point arithmetic often result in non-physical breaking of degeneracies. To resolve this issue, nearly degenerate states within a finite (small) energy window $\delta E$ are grouped and treated as degenerate. Users have an option to select the parameter $\delta E$, otherwise a default value of $10^{-6}$~Ha is implied. This approach is numerically stable as long as the group of nearly degenerate states is well separated in energy from other bands, \textit{i.e.}, when the condition $|E_D - E_l| \gg \delta E$ is fulfilled  in Eq.~(\ref{Eq:Mnm}).

\subsection{WIEN2k implementation details}

WIEN2k \cite{Blaha_WIEN2k_2018,Blaha_JCP_152_2020} is an all-electron implementation of DFT \cite{Hohenberg_PR_136_1964,Kohn_PR_140_1965} based on the augmented plane wave plus local orbitals method \cite{Singh_Planewaves_2006}. The main advantage of working with a full-potential code is the possibility to check the method in a framework that is free from  additional approximations, such as an effective potential for valence electrons, which is particularly important when testing new DFT functionals for which optimized pseudopotentials may not exist.\cite{BorlidoJCTC20}

To determine $m^*$, one would typically start with a self-consistent field (SCF) calculation to generate the charge density. Then, eigenvalues and wave functions need to be obtained for $k$ points of interest. It is important to include a large number of empty states. This is achieved by increasing an $E_\text{max}$ parameter in  \texttt{case.in1(c)} and \texttt{case.inso} input files up to at least 5~Ry (see Sec.~\ref{Sec:Method:Sample work flow} for details on the work flow).

Momentum matrix elements $p_{nl}^{(\alpha)}$ are computed using the optic module \cite{Ambrosch-Draxl_CPC_175_2006} of WIEN2k. Its input file \texttt{case.inop} is edited to enable writing of momentum matrix elements by switching to the option ``\texttt{ON}". The  $E_\text{max}$ parameter in the same file has to be adjusted to match the value set during SCF calculation. The  momentum matrix elements are tabulated in a formatted \texttt{case.mommat2(up/dn)} file in the following order for each $k$ point
\vspace{12pt}\\
	\indent\texttt{n~~l~~Re($p_{nl}^{(x)}$)~~Im($p_{nl}^{(x)}$)~~Re($p_{nl}^{(y)}$)~~Im($p_{nl}^{(y)}$)~~Re($p_{nl}^{(z)}$)~~Im($p_{nl}^{(z)}$)~~$\Delta E_{nl}$}
\vspace{12pt}\\
Here $n$ and $l$ are band indexes ($l \ge n$), $p_{nl}^{(\alpha)}$ are matrix elements in atomic units, $\Delta E_{nl}=E_l - E_n$ is the energy difference in Ry. This is the only input file required by \texttt{mstar}. It should be noted that the optic module in WIEN2k also generates a \texttt{case.symmat(up/dn)} file with squared matrix elements, which are used to calculate the imaginary part of the dielectric function.

The momentum matrix elements $p_{nl}^{(\alpha)}$  in WIEN2k are computed using the wave functions, i.e., $\left\langle \psi_{n} | \hat{p}_\alpha | \psi_{l} \right\rangle$, rather than the cell periodic part $u_{n\bm{k}}(\bm{r})$ [see the note below Eq.~(\ref{Eq:Hnm-k.p})]. However, it does not pose any difficulty in view of the relation
\begin{equation}\label{Eq:p-ME-with-psi}
	\langle u_{n\bm{k}_0} | (\hat{p}_\alpha + \hbar k_0^{(\alpha)} ) | u_{l\bm{k}_0} \rangle
	\equiv
	\left\langle \psi_{n\bm{k}_0} | \hat{p}_\alpha | \psi_{l\bm{k}_0} \right\rangle,
	\quad\quad
	n \neq l
\end{equation}
where the $\langle u_{n\bm{k}_0} |\hbar k_0^{(\alpha)}  | u_{l\bm{k}_0} \rangle$ term vanishes for off-diagonal matrix elements $(n \neq l)$. The diagonal matrix elements $p_{nn\bm{k}_0}^{(\alpha)}$ are present in Eq.~(\ref{Eq:Hnm-k.p}) for completeness, but they do not propagate into the effective mass calculation [see Eqs.~(\ref{Eq:eff-mass-nondegenerate}) and (\ref{Eq:Mnm})].

Calculation of effective masses is done by invoking \texttt{mstar}. Its execution involves two optional arguments: the spin channel and the degeneracy energy tolerance $\delta E$ (Ha units)
\vspace{12pt}\\
	\indent\texttt{x mstar [-up/-dn] [-settol 1.0e-5]}
\vspace{12pt}\\
The tolerance is optional with the default value of $\delta E=10^{-6}$~Ha. The code generates four output files: \texttt{minv\_ij.dat}, \texttt{minv\_pr.dat}, \texttt{minv\_c.dat}, and  \texttt{minv\_d.dat}. The files contain components of the inverse effective mass tensor $(m^*_{\alpha\beta,n}(\bm{k}))^{-1}$, principal components of the inverse effective mass tensor [eigenvalues of $(m^*_{\alpha\beta,n}(\bm{k}))^{-1}$], the inverse conductivity $(m^*_{\text{c},n}(\bm{k}))^{-1}$ and density of state $(m^*_{\text{d},n}(\bm{k}))^{-1}$ effective masses in units of $m_0$, respectively. The conductivity effective mass is defined as
\begin{equation}\label{Eq:m_c}
	(m^*_{\text{c}})^{-1} =
	\frac{1}{3} \, \text{Tr}[(m^*_{\alpha\beta})^{-1}]
	,
\end{equation}
while the density of states mass is expressed as
\begin{equation}\label{Eq:m_d}
	m^*_{\text{d}} =
	\sqrt[3]{m^*_1 \, m^*_2 \, m^*_3}
	,
\end{equation}
where $m^*_1$, $m^*_2$, and $m^*_3$ are the \textit{principal} components of the effective mass tensor. It should be noted that the trace of a tensor is invariant under axis rotation.

A sample listing of the output file \texttt{minv\_ij.dat} generated by \texttt{mstar} for GaAs with SOC and \citet*{Perdew_PRL_77_1996} (PBE) exchange-correlation (XC) approximation is shown below. The $k$ point 13 corresponds to $\Gamma$. A total of 138 bands were included in the calculation. Bands 1--28 are valence states (23--24, 25--26, and 27--28 are split-off (SO), light-hole (LH) and heavy-hole (HH) bands, respectively). Effective masses are double degenerate due to SOC and $\Gamma$ symmetry. At the  valence band maximum (VBM), the masses along Cartesian $[100]$ direction are $m^*_\text{SO}= (-9.640)^{-1} m_0 =-0.10m_0$, $m^*_\text{LH}=-0.034m_0$, and $m^*_\text{HH}=-0.32m_0$. Holes have a negative effective mass as the curvature of a parabola points downward. The conduction band minimum (CBM) has a positive effective mass $m^*_\text{CB}=0.027m_0$.
\begin{verbatim}
# This file is generated by mstar
# the output contains inverse effective masses (m0/m_ij*) that
# are grouped by k-point index and then by the band index
# columns correspond to Cartesian directions for m_ij
# band 1=xx;     2=yy;      3=zz;     4=yz;       5=xz;      6=xy
...
# KP: 13 NEMAX: 138
   1  8.741E-01  8.741E-01  8.741E-01 -1.950E-02 -1.950E-02 -1.950E-02
   2  8.741E-01  8.741E-01  8.741E-01 -1.950E-02 -1.950E-02 -1.950E-02
...
  21  8.345E-01  8.345E-01  8.345E-01  1.067E-08  1.627E-09 -6.878E-08
  22  8.345E-01  8.345E-01  8.345E-01  4.870E-08  1.513E-07  1.208E-07
  23 -9.640E+00 -9.640E+00 -9.640E+00  7.737E-07 -1.455E-05 -1.222E-05
  24 -9.640E+00 -9.640E+00 -9.640E+00  1.359E-05  1.061E-05 -6.368E-06
  25 -2.957E+01 -2.957E+01 -2.957E+01 -1.309E+01 -1.309E+01 -1.309E+01
  26 -2.957E+01 -2.957E+01 -2.957E+01 -1.309E+01 -1.309E+01 -1.309E+01
  27 -3.083E+00 -3.083E+00 -3.083E+00  1.309E+01  1.309E+01  1.309E+01
  28 -3.083E+00 -3.083E+00 -3.083E+00  1.309E+01  1.309E+01  1.309E+01
  29  3.648E+01  3.648E+01  3.648E+01  1.170E-05 -2.977E-05 -1.167E-05
  30  3.648E+01  3.648E+01  3.648E+01  2.271E-05  3.973E-05  1.813E-05
...
\end{verbatim}
The absence of off-diagonal components of the effective mass tensor for CB and SO (last three columns for bands 23--24, 29--30) combined with equal diagonal components indicate that the mass is isotropic. The latter agree with $s$-character of those states. Bands 25--28 have non-zero off-diagonal components, which indicates that their mass is anisotropic and principal axes are not aligned with  Cartesian coordinates. These bands have a dominant $p$-character. This analysis is corroborated by principal components (file \texttt{minv\_pr.dat}) listed below.
\begin{verbatim}
# This file is generated by mstar
# the output contains principal components of the inverse
# eff. mass tensor eig(m0/m*_ij) that
# are grouped by k-point index and then by the band index
# columns correspond to
# band m0/m_1  m0/m_2  m0/m_3
...
# KP: 13 NEMAX: 138
   1  8.351E-01  8.936E-01  8.936E-01
   2  8.351E-01  8.936E-01  8.936E-01
...
  21  8.345E-01  8.345E-01  8.345E-01
  22  8.345E-01  8.345E-01  8.345E-01
  23 -9.640E+00 -9.640E+00 -9.640E+00
  24 -9.640E+00 -9.640E+00 -9.640E+00
  25 -5.575E+01 -1.649E+01 -1.649E+01
  26 -5.575E+01 -1.649E+01 -1.649E+01
  27 -1.617E+01 -1.617E+01  2.309E+01
  28 -1.617E+01 -1.617E+01  2.309E+01
  29  3.648E+01  3.648E+01  3.648E+01
  30  3.648E+01  3.648E+01  3.648E+01
...
\end{verbatim}

\subsection{Sample work flow}\label{Sec:Method:Sample work flow}

We present step-by-step instructions on how to perform a calculation of effective masses for Kohn-Sham eigenstates in GaAs at the PBE level. The instructions include optional steps that extend this capability to an arbitrary $k$ path. Here we show an example with SOC, since it is essential for masses in GaAs. However, it is of course not mandatory to use SOC for all materials, and corresponding parts can be skipped. (WIEN2k version 20.1 was used.)

\begin{itemize}
	\item Generate structure file using \texttt{w2web} or the \texttt{makestruct} utility: F-type cubic lattice, $a=5.653$~{\AA}, 2 atoms (Ga and As) with coordinates $(0,0,0)$ and $(0.25,0.25,0.25)$, respectively.
	\item Initialize calculation with PBE XC, $R_{\text{MT}}^\text{min}K_\text{max}=7$, and 500 $k$ points in the full Brillouin zone\\
		\texttt{init\_lapw -b -vxc 13 -rkmax 7 -numk 500}
	\item Run SCF cycle with the energy convergence of $10^{-5}$~Ry and the charge convergence of $10^{-4}e$\\
		\texttt{run\_lapw -ec 0.00001 -cc 0.0001}
	\item Save calculation\\
		\texttt{save\_lapw -d noSOC}
	\item Initialize SOC calculation without relativistic $p_{1/2}$ local orbitals (since they are not supported in optic) \\
		\texttt{init\_so\_lapw}
	\item Run SCF cycle with SOC\\
		\texttt{run\_lapw -ec 0.00001 -cc 0.0001 -so}
	\item Increase the number of empty states by setting $E_\text{max}=5$~Ry in \texttt{case.in1c} and \texttt{case.inso} input files
	\item (optional) Generate $L-\Gamma-X$ $k$ path with 100 intermediate points using \texttt{xcrysden} \cite{Kokalj_CMS_28_2003} and save the path as \texttt{case.klist\_band}
	\item Generate eigenvalues and wave functions (including high energy ones) for $k$ point (add \texttt{-band} option only if $k$ points from the \texttt{case.klist\_band} file should be used)\\
		\texttt{x lapw1 [-band]}\\
		\texttt{x lapwso}
	\item Create symbolic links (or copy) to emulate a spin-polarized calculation for the optics module (these links should be removed after an effective mass calculation and before other calculations can be performed)\\
		\texttt{ln -s case.vsp case.vspup}\\
		\texttt{ln -s case.vsp case.vspdn}\\
		\texttt{ln -s case.vectorso case.vectorsoup}
	\item Copy the optic input file and enable writing of matrix elements (change writing option to ``\texttt{ON}") as well as extend $E_\text{max}$ to 5~Ry in\\
		\texttt{cp \$WIENROOT/SRC\_templates/case.inop case.inop}
	\item Calculate optical matrix elements\\
		\texttt{x optic -so -up}
	\item Calculate effective masses with the degeneracy energy tolerance parameter $\delta E=10^{-5}$~Ha\\
		\texttt{x mstar -up -settol 1.0e-5}
\end{itemize}

\section{Results and Discussion}

\subsection{Validation of numerical results}

Validation of results for effective masses obtained using PT [Eqs.~(\ref{Eq:Mnm}) and (\ref{Eq:eff-mass-degenerate})] will be performed by comparing with results obtained from numerical differentiation of the band dispersion. For this purpose, we selected three well-characterized solar cell materials: Si (diamond structure), GaAs (zinc blende structure), and CdTe (zinc blende structure). Calculations were performed at experimental lattice parameters\cite{Landolt-Boernstein_1972_vol_III}: $a_0=5.431$~{\AA} for Si, 5.653~{\AA} for GaAs, and 6.48~{\AA} for CdTe. Two exchange-correlation approximations were used: PBE and the Tran-Blaha modified Becke-Johnson potential (TB-mBJ) \cite{Tran_PRL_102_2009}. The last  approximation was selected due to its more accurate predictions for the band gap as compared to PBE. SOC was included for all compounds. The band curvature $\partial^2 E_{n}/\partial k_{\alpha}\partial k_{\beta}$ was extracted from a band dispersion by fitting $E_{n}(q)$ to a 4th order polynomial function in the vicinity of a band extremum at $\bm{k}_0$ [similar to Eq.~(\ref{Eq:E(q)-second-order-q})] within the energy window of 20--30~meV using at least 7 $k$ points. The higher order terms in the polynomial account for non-parabolicity of bands, which is particularly important for the LH band and the conduction band of GaAs.

Effective masses listed in Table~\ref{tab:eff-mass} show a quantitative agreement between the band curvature (numerator) and PT (denominator) data for both PBE and TB-mBJ XC approximations (agreement with experimental values is not essential at this point). PBE masses are generally lighter than experimental values due to the severe underestimation of the band gap. TB-mBJ recovers the band gap error resulting in effective masses becoming more consistent with the experiment, albeit being on the heavy side as noted earlier \cite{Kim_PRB_82_2010}.

Results for $m^*$ obtained in the framework of PT are sensitive to momentum matrix elements that involve upper energy bands. Thus, it is important to check the convergence with respect to the number of bands. Our experience shows that least dispersive bands are more difficult to converge. For instance, this is the case for the conduction band edge of Si (the longitudinal mass) in Table~\ref{tab:eff-mass}. Figure~\ref{Fig:Si-convergence} shows sensitivity of $m^*$ in Si to the number of bands included in the perturbation sum [Eq.~(\ref{Eq:Mnm})]. The magnitude of $m_\text{e}^{\text{long}}$ makes an abrupt change near 30--50 bands, after which $m_\text{e}^{\text{long}}$ converges very slowly towards its asymptotic value given by the band curvature. Data reported in Table~\ref{tab:eff-mass} are obtained with $E_\text{max}=5$~Ry, which corresponds to approximately 120, 140, and 200 bands for Si, GaAs, and CdTe, respectively.

\subsection{Illustrative applications}

\subsubsection{Graphene}

Graphene is a 2D semi-metal with peculiar electronic properties. Its  low-energy charge carriers exhibit a linear dispersion relation $E=\hbar v_\text{F} k_\text{F}$ inherent to ultrarelativistic particles in spite of the fact that their group velocity $v_\text{F}$ is much less than the speed of light $c$ \cite{Neto_RMP_81_2009}.  Here the Fermi wave vector $\bm{k}_\text{F} = \bm{k} - \bm{K}$ is  defined relative to the Dirac point $K$ in the Brillouin zone where the band crossing occurs (Fig.~\ref{Fig:Graphene}a,b). The first experimental evidence of a linear dispersion relation in graphene came from measurements of the cyclotron effective mass  \cite{Novoselov_N_438_2005,Zhang_N_438_2005}. 

In graphene, there are two principal components (perpendicular and parallel to $\bm{k}_\text{F}$) of the effective mass tensor. They show a strong directional dependence as can be inferred from Fig.~\ref{Fig:Graphene}c,d. (It should be emphasised that with the PT-based method the effective mass distribution can be obtained conveniently by using only $k$ points located on the high-symmetry path, which would be insufficient to determine the band curvature using a finite difference.) The lightest mass corresponds to a perpendicular component as shown schematically in Fig.~\ref{Fig:Graphene}b; the parallel mass is much heavier ($m^* \sim m_0$) because of the nearly linear band dispersion. The remaining discussion is focused on the perpendicular mass.

The electron mass becomes progressively lighter as its wave vector approaches the Dirac point. In the vicinity of a Dirac crossing, the effective mass is expected to become proportional to $k_\text{F}$ \cite{Neto_RMP_81_2009}
\begin{equation}\label{Eq:m*-linear}
	m^* = \hbar k_\text{F}/v_\text{F}.
\end{equation}
This relationship is also observed in our calculations (Fig.~\ref{Fig:Graphene}e) allowing us to deduce the Fermi velocity $v_\text{F}=c\,(315-365)^{-1}$ from its slope, which is compatible with the experimental result \cite{Novoselov_N_438_2005} $v_\text{F} \approx c/300$.

In experiment, one manipulates $k_\text{F}$ indirectly by varying the carrier density and measuring the cyclotron effective mass as its function  \cite{Novoselov_N_438_2005,Zhang_N_438_2005}. In neutral (defect-free) graphene, the Fermi energy $E_\text{F}$ coincides exactly with the Dirac point $E_\text{D}$, and the mass approaches zero since $k_\text{F}=0$ . However, a finite carrier density $n_\text{e}$ leads to a finite $k_\text{F}$ and, thus, a finite mass. An experiment \cite{Novoselov_N_438_2005} yielded the cyclotron mass $m^*$ varying between 0.02 and $0.07m_0$  for the range of carrier concentration $n_e=(1-7)\times10^{12}$~cm$^{-2}$, which is in quantitative agreement with DFT calculations (Fig.~\ref{Fig:Graphene}f). It should be noted that the experimental cyclotron mass is not associated with any specific direction in the reciprocal space, but rather represents an average value \cite{Shockley_PR_90_1953}. Lighter mass ($\sim 0.01m_0$), and thus greater electron mobility, can only be achieved \cite{Zhang_N_438_2005,Tiras_JAP_113_2013} at a lower carrier density $n_\text{e} \sim 2\times10^{11}$~cm$^{-2}$. A square root dependence of the cyclotron mass on the electronic density \cite{Novoselov_N_438_2005,Zhang_N_438_2005}, that became a landmark of the Dirac-like dispersion, is reproduced in our calculations (Fig.~\ref{Fig:Graphene}f) using an \textit{ab initio} density of states. Calculations were performed at PBE level using an experimental  lattice parameter of 2.46~{\AA} and a vacuum thickness of 20~bohrs.

\subsubsection{CuI: Cu-vacancy and alloyed with Sn}

CuI (zinc blende structure) has emerged as a high-mobility p-type wide band-gap  semiconductor \cite{Chen_CGD_10_2010} offering one of the best combinations of conductivity and transparency to visible light \cite{Yang_PNAS_113_2016} among existing p-type transparent conducting materials. The high mobility of holes (44~\ch{cm^2~V^{-1}~s^{-1}}, Ref.~\citenum{Chen_CGD_10_2010}) is attributed to a low effective mass of a light hole band \cite{Ferhat_MSEB_39_1996,Huang_JPDAP_45_2012}. The first requirement to quality effective mass calculations is an accurate band gap. The experimental band gap of cubic CuI is $3.0-3.05$~eV \cite{Cardona_PR_129_1963,Chen_CGD_10_2010}, but standard PBE calculations yield only $1.18$~eV\cite{Zhang_PRM_4_2020}. Our calculation yields the band gap of $2.96$~eV, which is achieved by employing the TB-mBJ XC potential with an effective Hubbard-like term \cite{Dudarev_PRB_57_1998} for the Cu-$d$ states of $U=0.36$~Ry without an on-site exchange ($J=0$).

The band structure of CuI is presented in Fig.~\ref{Fig:CuI}a. This compound presents a challenging case for the effective mass calculation, in particular for the valence bands, which have a strong contribution of fairly localized Cu-$d$ electrons. To achieve accurate effective masses from PT (\textit{i.e.}, those in agreement with the numerical band curvature), it is necessary to extend the basis set by including high energy local orbitals  (HELOs) for a better description of unoccupied states \cite{Laskowski_PRB_85_2012}. In the case of CuI, $s$-, $p$-, $d$-, and $f$-LOs are added. The necessity for HELOs can be rationalized via an  electric dipole selection rule $\Delta \ell \pm 1$ ($\ell$ is an azimuthal quantum number). Thus, strong momentum matrix elements in Eq.~(\ref{Eq:eff-mass-nondegenerate}) are expected between $d-f$ states and $p-d$ states. The convergence tests (Fig.~\ref{Fig:CuI}b) show that it is not only essential to add one HELO, but we need at least three HELOs per angular momentum $\ell$. It should be noted that in WIEN2k each subsequent HELO is automatically chosen such that it has an additional node in the corresponding radial function. The higher the number of nodes, the higher is the energy of the orbital. For instance, the third $\ell=3$ HELO has 3 nodes and corresponds to a $7f$ state positioned at 100~Ry above the Fermi energy. The mean absolute relative error in $m^*$ (the numerical band curvature \textit{vs} PT, Fig.~\ref{Fig:CuI}b) drops from 40\% (without HELOs) down to 6\% with 3 or more HELOs. (The HELO expansion of a basis set is enabled in a very limited number of DFT codes, which is another advantage of WIEN2k for testing the PT implementation).

However, in the case of Si, the expanded basis set with HELOs leads to an ``overcorrection" of the longitudinal effective mass in the conduction band ($m_\text{e}^{\text{long}}/m_0=1.01$). It is due to  neglect of matrix elements with core states. When Si-2$p$ semicore states are included as valence states, we obtain the most accurate result ($m_\text{e}^{\text{long}}/m_0=0.97$ with HELOs), which is only 1\% off the value derived from the band curvature (Table~\ref{tab:eff-mass}). A similar interplay between HELOs and semicore states was noted in calculations of a magnetic shielding for solid state nuclear magnetic resonance chemical shifts \cite{Laskowski_PRB_89_2014}.

Having established the basis set for effective mass calculations, we can now explore effects of structural defects on the electronic structure, and $m^*$ specifically, in view of its connection to mobility of charge carriers. Copper vacancies have the lowest formation energy among native defects in CuI \cite{Wang_JAP_110_2011} and are responsible for its p-type conductivity. The presence of Cu vacancies has a marginal effect on the band dispersion near to the band edges (compare Fig.~\ref{Fig:CuI-v_Cu}a with Fig.~\ref{Fig:CuI}a). The effective masses of holes become only $10-15$\% heavier. The tolerance of CuI to defects can be attributed to an antibonding nature of both CBE and VBE (Fig.~\ref{Fig:CuI-v_Cu}b) similar to halide perovskites \cite{Zheng_JCP_151_2019,Goesten_JACS_140_2018,Yin_APL_104_2014}.

\citet{Jun_AM_30_2018} suggested alloying of CuI with $5 - 10$~mol\% Sn to stabilize an amorphous phase without a significant penalty in p-type mobility relative to a polycrystalline CuI. Here we explore the effect of Sn incorporation on the electronic structure of CuI. In contrast to Ref. \onlinecite{Zhang_PRM_4_2020}, where the structure was modelled as an ultra-fast quenched amorphous state,  we model the defects by using $2 \times 2 \times 2$ supercells (64 atoms), where we substitute one Sn$^{4+}$ for 4 Cu$^{+}$ ions. The Sn content $x$ is defined as \ch{Cu_{$4(1-x$)}Sn_{$x$}I_{4}}, which represents an interpolation between two stoichiometric compounds CuI and \ch{SnI4} and should not be confused with mol\% Sn. Since the position of Sn atoms in CuI and its local coordination are unknown, several CuI:Sn models were created. Models with the lowest energy are shown in Fig.~\ref{Fig:CuI+Sn}a,b referred to as substitutional Sn with tetrahedral \ch{SnI4} and interstitial Sn with octahedral \ch{SnI6} coordination. The latter model is about 0.4~eV lower in energy.  

Incorporation of Sn in CuI ($x=12.5$\%) is accompanied by formation of localized states within the fundamental gap of the host CuI, which can be seen as non-dispersive lines in the band structure (Fig.~\ref{Fig:CuI+Sn}d,e). Effectively, the band gap shrinks in agreement with experimental observation \cite{Li_AMI_6_2019}. However, these localized states should not hinder p-conductivity, since they are well separated from the valence band. A disparity in the effective mass between light and heavy holes at the top of the valence band is reduced but remains comparable to CuI with Cu vacancies (Fig.~\ref{Fig:CuI-v_Cu}b). At a high concentration of Sn ($x=37.5$\%, Fig.~\ref{Fig:CuI+Sn}c), localized states also emerge at the top of the valence band (Fig.~\ref{Fig:CuI+Sn}f), which is consistent with a steep decline of p-type mobility observed experimentally at a higher Sn content past 1~mol\% \cite{Li_AMI_6_2019}, which corresponds to $x \sim 8$\% in our structures.

\subsection{Limitations}\label{Sec:Results:Limitations}

The first order term in the series expansion of the Hamiltonian of the cell-periodic wave function [Eq.~(\ref{Eq:H(k0+q)-k.p})] is, more generally, expressed as \cite{Boykin_PRB_52_1995} $\bm{q}\cdot \nabla_{\bm{k}_0} \hat{H}(\bm{k}_0)$. It is evaluated using a commutation relation
\begin{equation}\label{Eq:dH/dk}
	\frac{\partial \hat{H}}{\partial \bm{k}}
	= i[\hat{H},\bm{r}]
	= \frac{\hbar}{m_0} \left( \hat{\bm{p}} + \hbar\bm{k}_0 \right) + i [\hat{V},\bm{r}]
	,
\end{equation}
and is equivalent to the \textit{velocity} operator \cite{Starace_PRA_3_1971}. Here $\hat{V}$ is the potential operator. Local potentials, such as PBE or TB-mBJ, commute with the position operator, \textit{i.e.},  $[\hat{V},\bm{r}]=0$. In the case of non-local potentials, such as Hartree-Fock or hybrid, the potential no longer commutes with the position operator, and the $i[\hat{V},\bm{r}]$ term cannot be ignored \cite{Rhim_PRB_71_2005}. \citet{Pickard_PRB_62_2000} discussed this issue in the context of effective masses, whereas the importance of a velocity (rather than momentum) operator in calculations of optical properties with a non-local Hamiltonian has been emphasised a long time ago \cite{Starace_PRA_3_1971}.

At present, the term $i[\hat{V},\bm{r}]$ is not implemented in WIEN2k, which restricts our PT effective mass analysis (as well as the calculation of optical properties) to local potentials (and DFT+$U$, see below) only. We need to mention that, if momentum matrix elements were used in conjunction with the hybrid functional YS-PBE0\cite{Tran_PRB_83_2011}, PT would erroneously predict systematically heavier masses (25\% heavier on average). However, it is more common to account for the $i[\hat{V},\bm{r}]$  term in pseudopotential DFT codes \cite{Gajdos_PRB_73_2006}, where the non-locality is also a part of a pseudopotential itself. Actually, DFT+$U$ also leads to a non-local potential, however the type of non-locality is different (weaker) from the non-locality of Hartree-Fock/hybrid. We have not observed problems when using the PT method with DFT+$U$ and, thus, believe that the term $i[\hat{V},\bm{r}]$ is not required in this case. We expect a more general $\bm{k} \cdot \bm{v}$ formalism (see Ref.~\citenum{Rubel_git_mstar}) to be also compatible with non-local potentials, \textit{e.g.}, Hartree-Fock/hybrid.

\section{Conclusion}

A degenerate perturbation $\bm{k}\cdot\bm{p}$ approach for effective mass calculations has been implemented in the all-electron DFT package WIEN2k. It essentially yields the same results as the band curvature when applied to major group IVA,  IIIA-VA, and IIB-VIA semiconductor materials with $sp^3$ bonding, provided momentum matrix element with sufficiently high-energy states (about 5~Ry above the Fermi energy) are included. For accurate comparison of effective masses with experiment, it is essential to select an exchange-correlation approximation that reproduces the band gap. A quantitative agreement between experiment and theory is demonstrated for the effective mass as a function of the carrier density in graphene.

It is more challenging to apply the $\bm{k} \cdot \bm{p}$ formalism to systems where $d$-electrons contribute to states of interest, such as the valence band of CuI. For those states, the effective mass converges very slowly indicating the involvement of very high-energy states. High-energy local orbitals offer an efficient way to circumvent this issue. However, calculations become computationally more demanding, since all eigenvalues need to be computed. The analysis of CuI shows that its electronic structure is immune to defects (Cu vacancies). Incorporation of Sn  in quantities about 13\% as a stabilizer for an amorphous phase causes no harm to the effective mass in the valence band.

It is possible to extend application of the $\bm{k} \cdot \bm{p}$ formalism beyond the density functional theory, \textit{e.g.}, to hybrid calculations. Here we need to acknowledge a difference between the momentum and velocity operators in the context of non-local potentials. Calculations of effective masses would be still possible, provided \textit{velocity}  matrix elements are used.

\begin{acknowledgements}
O.R. acknowledges funding provided by Natural Sciences and Engineering Research Council of Canada (NSERC) under the Discovery Grant Program RGPIN-2020-04788.  Calculations were performed using a Compute Canada infrastructure supported by the Canada Foundation for Innovation under the John R. Evans Leaders Fund program and supercomputer resources at the Vienna Scientific Cluster. X.R. would like to acknowledge access to the HPC resources of [TGCC/CINES/IDRIS] under allocation 2019-A0010907682 made by GENCI.
\end{acknowledgements}

\bibliography{bibliography}

\clearpage

\begin{table}
\caption{\label{tab:eff-mass}Effective masses (units of $m_0$) at the conduction and valence band extrema of Si, GaAs, and CdTe calculated by evaluating a local band curvature using Eq.~(\ref{Eq:eff-mass-curvature}) (numerator) \textit{vs} the second order perturbation theory (denominator). Two options for the XC functional/potential were selected that yield a different accuracy for the band gap $E_\text{g}$. Experimental values of the fundamental band gap are listed at $T \rightarrow 0$~K. The negative sign in $m$ for the valence band is omitted.}
\begin{ruledtabular}
\begin{tabular}{lcccccc}
  Method & $m_{\text{LH}}^{\langle 100\rangle}$ & $m_{\text{HH}}^{\langle 100\rangle}$ & $m_{\text{SO}}$\footnote{At $\Gamma$ point, isotropic.} & $m_\text{e}^{\text{long}}$  & $m_\text{e}^{\text{trans}}$ & $E_\text{g}$~(eV)\\
  \hline
   Si\\
   ~~PBE & 0.19/0.19 & 0.26/0.27 & 0.23/0.23 & 0.96/0.93\footnote{At $\Delta$ point in direction $[100]$.} & 0.20/0.19\footnote{At $\Delta$ point in direction $[010]$.} & 0.56\\
   ~~TB-mBJ  &  0.24/0.24 & 0.32/0.34 & 0.28/0.29 &  0.96/0.94 & 0.22/0.21 & 1.15\\
   ~~experiment &  0.18\cite{Ramos_PRB_63_2001} & 0.46\cite{Dexter_PR_96_1954} & 0.23\cite{Ramos_PRB_63_2001} & 0.92\cite{Ramos_PRB_63_2001} & 0.19\cite{Ramos_PRB_63_2001} & 1.17\cite{Bludau_JAP_45_1974}\\
   \hline
   GaAs\\
   ~~PBE & 0.033/0.034 & 0.32/0.32 & 0.11/0.10 & \multicolumn{2}{c}{0.027/0.027\textsuperscript{a}} & 0.40\\
   ~~TB-mBJ & 0.11/0.11 & 0.36/0.37 & 0.21/0.20 & \multicolumn{2}{c}{0.090/0.090} & 1.55\\
   ~~experiment & 0.085\cite{Walton_JPCSSP_1_1968} & 0.34\cite{Nakwaski_PB_210_1995} & 0.17\cite{Vurgaftman_JAP_89_2001} & \multicolumn{2}{c}{0.067\cite{Nakwaski_PB_210_1995}} & 1.52\cite{Vurgaftman_JAP_89_2001}\\
   \hline
   CdTe\\
   ~~PBE & 0.054/0.057 & 0.45/0.45 &  0.25/0.25 & \multicolumn{2}{c}{0.048/0.049\textsuperscript{a}} & 0.46\\
   ~~TB-mBJ & 0.15/0.16 & 0.52/0.56 & 0.36/0.37 &  \multicolumn{2}{c}{0.13/0.13} & 1.47\\
   ~~experiment & 0.12\cite{Madelung2004} & 0.53\cite{Dornhaus_properties_1976} & --- &  \multicolumn{2}{c}{0.11\cite{Marple_PR_129_1963}} & 1.60\cite{Madelung2004}\\
\end{tabular}
\end{ruledtabular}
\end{table}

\clearpage
\begin{figure}[h]
  \includegraphics{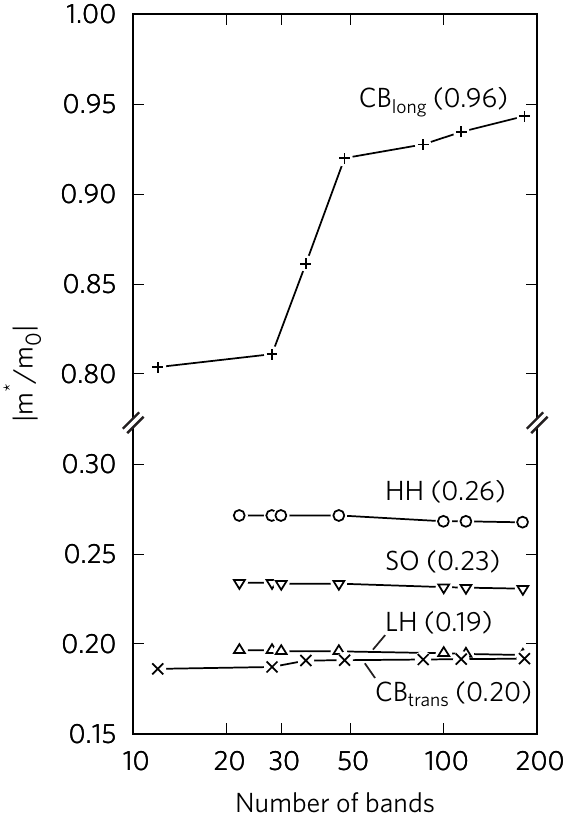}
  \caption{}
  \label{Fig:Si-convergence}
\end{figure}


\clearpage
\begin{figure}[h]
  \includegraphics[width=\textwidth]{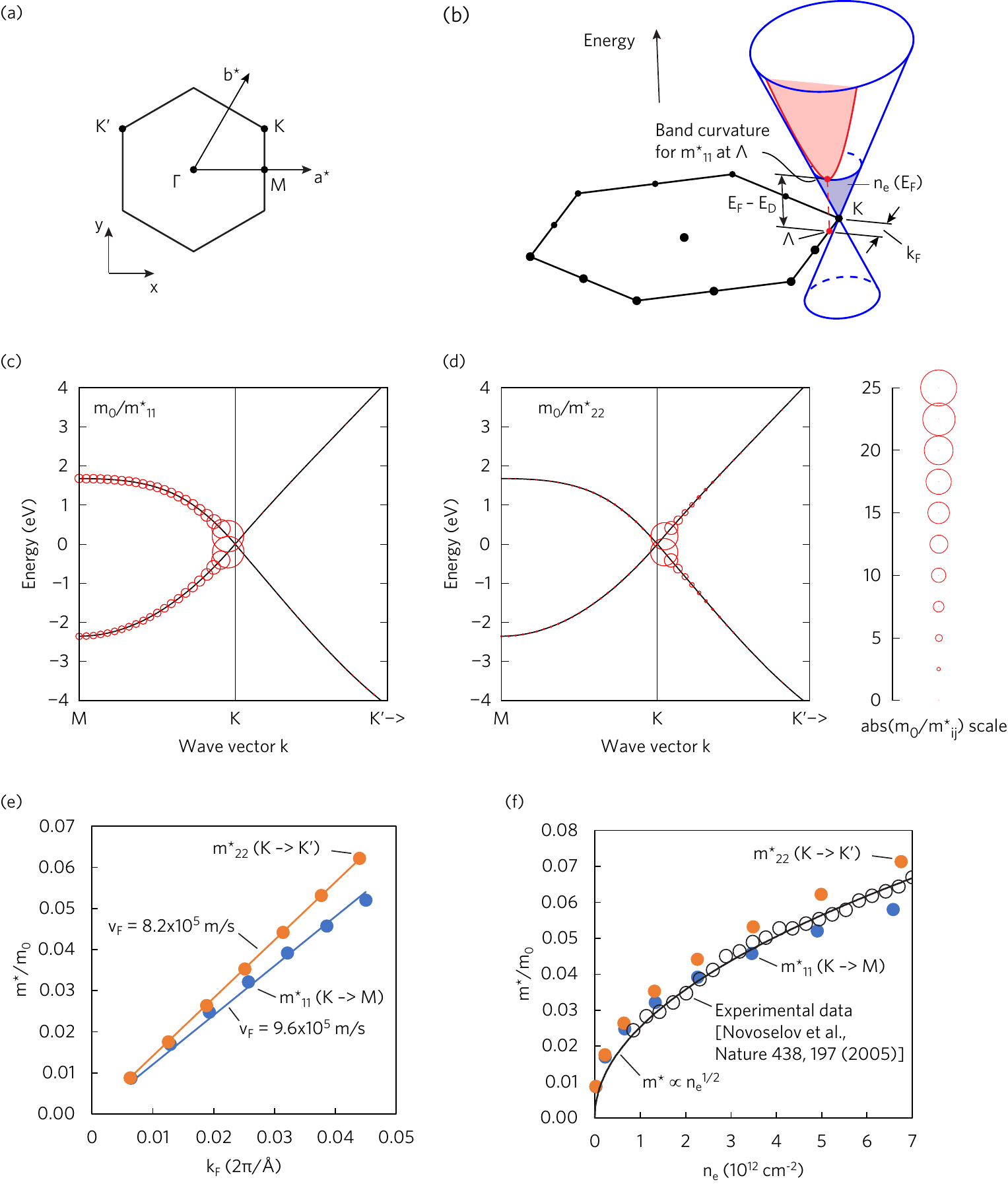}
  \caption{}
  \label{Fig:Graphene}
\end{figure}

\clearpage
\begin{figure}[h]
  \includegraphics{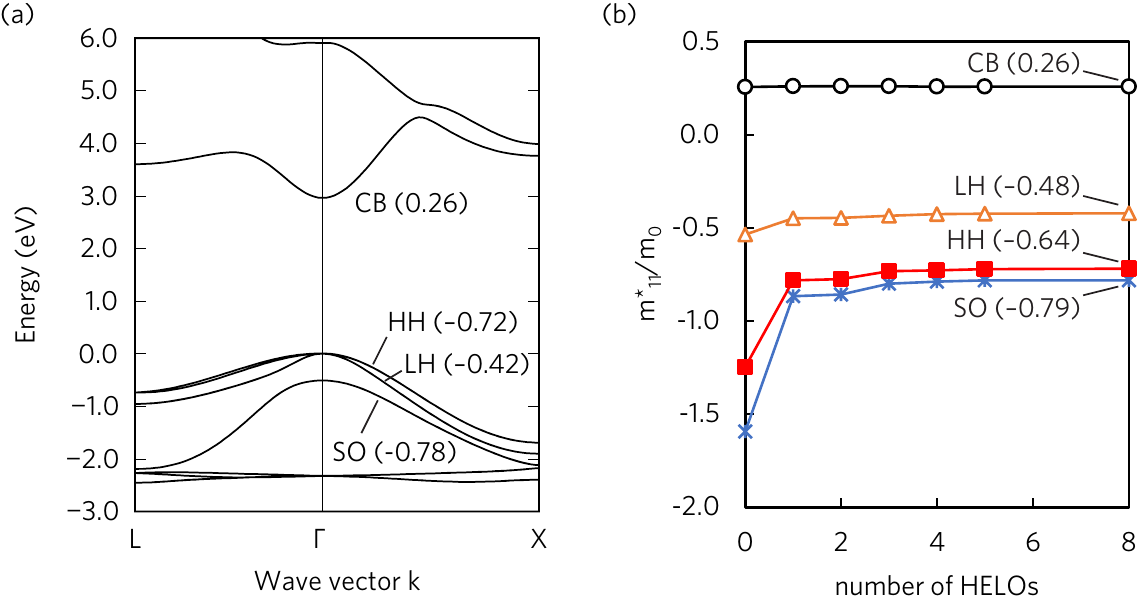}
  \caption{}
  \label{Fig:CuI}
\end{figure}

\clearpage
\begin{figure}[h]
  \includegraphics{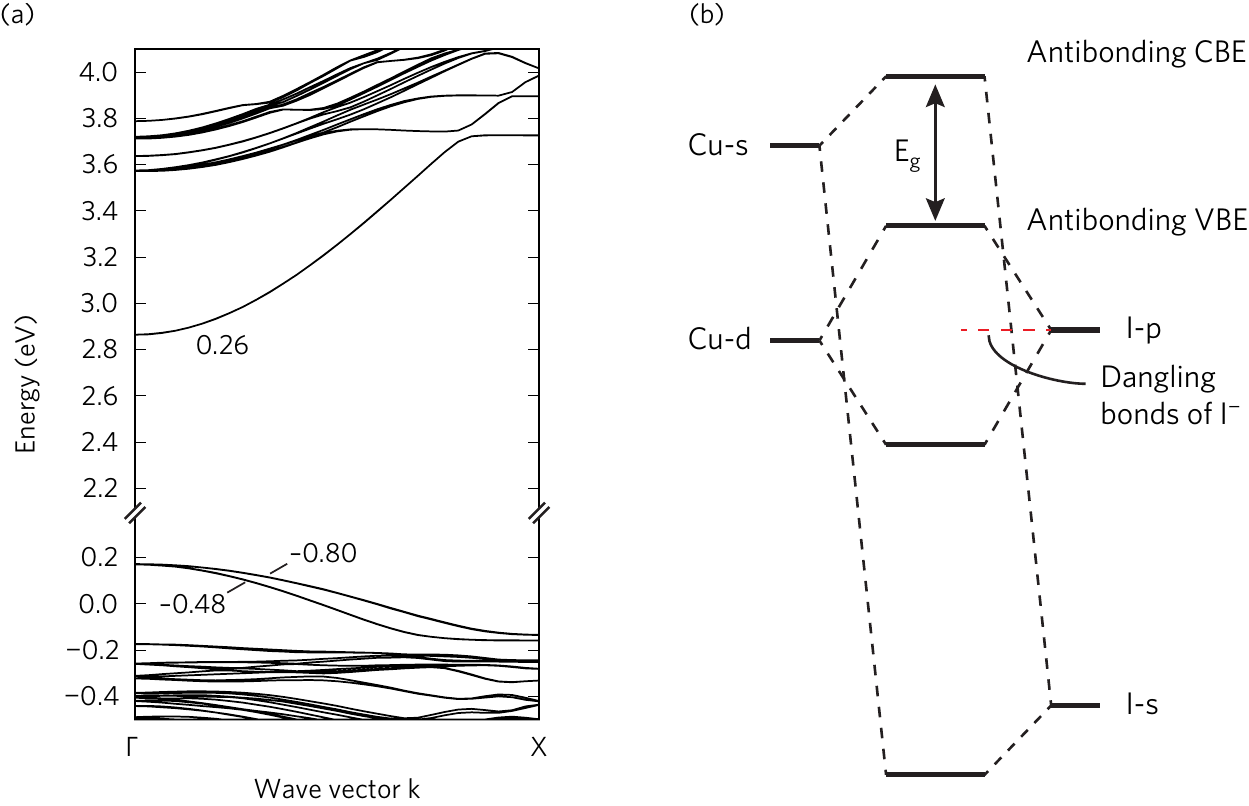}
  \caption{}
  \label{Fig:CuI-v_Cu}
\end{figure}

\clearpage
\begin{figure}[h]
  \includegraphics{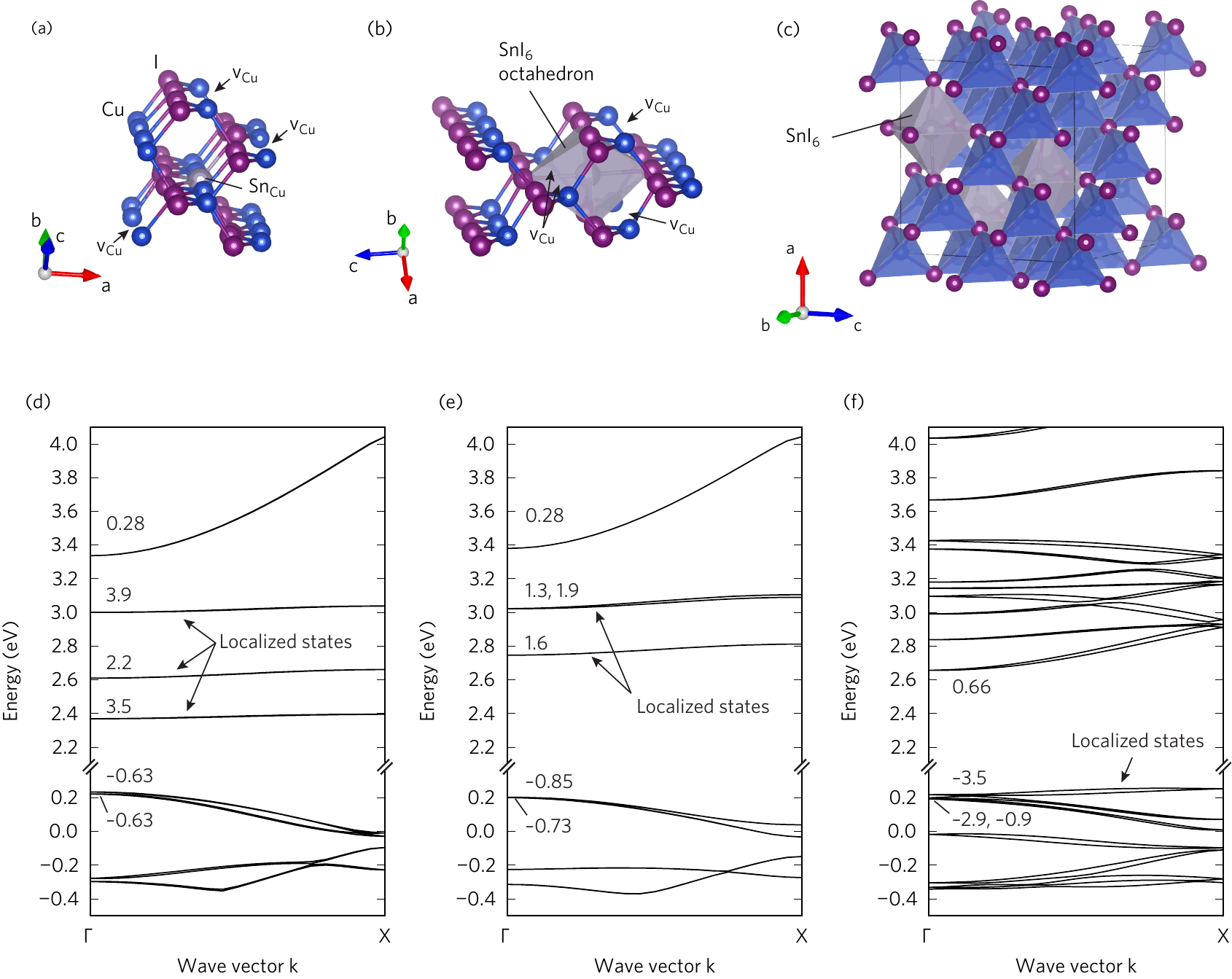}
  \caption{}
  \label{Fig:CuI+Sn}
\end{figure}

\clearpage

\textbf{Figure captions}:
\vspace{18pt}

\noindent\begin{minipage}{\textwidth}
FIG.~\ref{Fig:Si-convergence}: Convergence test for $m^*$ in Si (calculated at the PBE level) with respect to the number of bands (8 occupied bands) included in Eq.~(\ref{Eq:Mnm}). Values of $|m^*/m_0|$ extracted from the band curvature are listed in brackets. Accurate calculation of the longitudinal effective mass in the conduction band requires a large number of bands. With 180 bands the residual error drops below 2\%.
\end{minipage}\vspace{18pt}

\noindent\begin{minipage}{\textwidth}
\noindent FIG.~\ref{Fig:Graphene}: Effective mass in a monolayer graphene. (a) Brillouin zone in Cartesian coordinates. (b) Schematic energy band dispersion in the vicinity of the Dirac point $E_\text{D}$. (c,d) Cartesian components of the inverse effective mass tensor laid over the band structure calculated at the PBE level. The origin of the energy scale is set at the highest occupied eigenstate. (e) Fermi velocity $v_\text{F}$ deduced from the slope of a linear relationship between the effective mass and the Fermi wave vector $k_\text{F}$. (f) Effective mass as a function of the carrier density.  Filled markers represent DFT values, open circled correspond to the cyclotron effective mass measured experimentally \cite{Novoselov_N_438_2005}. The solid line shows a square root dependence of the mass on the carrier density \cite{Novoselov_N_438_2005}.
\end{minipage}\vspace{18pt}

\noindent\begin{minipage}{\textwidth}
FIG.~\ref{Fig:CuI}: (a) Band structure of CuI calculated at the TB-mBJ+$U$ level.  The origin of the energy scale is set at the highest occupied eigenstate. Values refer to the [100] effective masses at the $\Gamma$ point calculated using PT. (b) Convergence of $m^*/m_0$ at the $\Gamma$ point ([100] Cartesian component) with respect to the number of high energy local orbitals (HELOs). Numerical labels in brackets correspond to $m^*/m_0$ obtained from the band curvature.
\end{minipage}\vspace{18pt}

\noindent\begin{minipage}{\textwidth}
FIG.~\ref{Fig:CuI-v_Cu}: (a) Band dispersion along [100] direction in a \ch{Cu26I27} supercell with a Cu vacancy (\ch{v_{Cu}}) calculated at the TB-mBJ+$U$ level. Effective masses change only marginally relative to the defect-free material. The Fermi energy ($E=0$) is located below the VBE of CuI leading to an effective carrier density of $n_p=7\times 10^{20}$~cm$^{-3}$.  (b) Orbital energy diagram illustrating the tolerance of CuI to \ch{v_{Cu}} defects. Defect states due to dangling \ch{I-} bonds are expected \textit{within} the valence band away from the fundamental band gap.
\end{minipage}\vspace{18pt}

\newpage

\noindent\begin{minipage}{\textwidth}
	\noindent FIG.~\ref{Fig:CuI+Sn}: Incorporation of Sn into the CuI lattice and its effect on the electronic structure: (a) Local atomic configuration of a substitutional \ch{Sn_{Cu}} defect with a tetrahedral \ch{SnI4} coordination. 3 additional Cu vacancies (\ch{v_{Cu}})  are created to balance the formal charges of Sn$^{4+}$ and Cu$^{+}$. (b) Interstitial position of Sn within an octahedral \ch{SnI6} coordination and 4 nearest neughbor Cu$^{+}$ vacancies. (c) Supercell with 3 \ch{SnI6} octahedra. (d--f) Band dispersion along [100] direction in CuI with Sn defects at a substitutional site (\ch{Cu_{$4(1-x$)}Sn_{$x$}I_{4}}, $x=12.5$\%), an octahedral interstital site ($x=12.5$\%), and three octahedral defects ($x=37.5$\%), respectively. The origin of the energy scale is set at the highest occupied eigenstate. The origin of the energy scale is set at the highest occupied eigenstate.  Numerical labels on panels (d--f) correspond to the effective mass $m^*/m_0$ in $[100]$ direction at $\Gamma$ point. Defects give rise to localized states within the fundamental band gap.
\end{minipage}

\end{document}